\documentclass[aps,prx,reprint,groupedaddress]{revtex4-2}
\usepackage{amsmath}
\usepackage{amssymb}
\usepackage{graphicx}
\usepackage{epstopdf}
\usepackage{caption}
\usepackage[colorlinks,linkcolor=blue,urlcolor=blue,anchorcolor=blue,citecolor=blue]{hyperref}

\begin{document}
\title{Mechanical Oscillator Can Excite an Atom Through the Quantum Vacuum}
\author{Miao Yin}
\email{Corresponding author: scmyin@scut.edu.cn}
\affiliation{School of Physics and Optoelectronics, South China University of Technology, Guangzhou 510641, China}
\date{\today}
\begin{abstract}
We consider a two-photon Rabi model with one of the cavity mirrors connected by a mechanical oscillator in strong-coupling regime. We find that when the cavity is in its vacuum state, there exists a resonant coupling between the atom and mechanical oscillator even if the quality factor of the cavity is ultra low. The coupling is coherent and can be achieved by the exchange of virtual photon pairs induced by dynamical Casimir effect. Moreover, when considering the one-photon Rabi model, we find that the atom can absorb one photon from a virtual photon pair, leaving the other converting to a real photon. The behavior shows analogy with the well-known Hawking radiation. The parameters used in our theoretical models are all feasible data in experiments at present. Our theory reveals a kind of novel effective interaction and may find applications ranging from quantum information to nanotechnology.
\end{abstract}
\maketitle
\section{Introduction} \label{sec::1}
The interaction between light and matter is always one of the hot topics in physics. The simplest model describes the interaction is the Rabi model\cite{Rabimodel}, which shows a direct coupling between a two-level system (atom) and a bosonic mode (photon). In Rabi model, the photon number and atomic occupation show alternating cosinoidal oscillations with a certain frequency $\Omega_R$, termed as Rabi oscillation. It shows a coherent coupling and energy exchange between light and matters and represents an important theoretical building block of quantized matter-field interactions.
\par When the coupling strength between light and atom is weak, the rotating-wave approximation (RWA) can be applied, resulting a simplified Jaynes-Cummings (JC) model\cite{JC}. However, when the coupling strength exceeds the the losses of the system and even becomes comparable with bare frequencies of the uncoupled systems, the strong-coupling(SC) regime, ultrastrong-coupling (USC) regime, and even deepstrong-coupling (DSC) regime\cite{RevModPhys.91.025005} can be reached. The SC regime is first observed in Rydberg atoms confined in a high-Q cavity\cite{PhysRevLett.51.1175} and also experimentally confirmed in several physical systems\cite{PhysRevLett.68.1132,RevModPhys.85.1083}. In SC regime and USC regime, the RWA ceases to be applicable and the counter-rotating (number nonconserving) terms should be took into consideration and these terms can bring plenty of novel quantum effects\cite{RevModPhys.91.025005}. Photon blockade is revealed in a coherently driven Rabi model\cite{PhysRevLett.109.193602} and multiphoton blockade is recently reported in a two-photon JC model\cite{PhysRevA.102.053710}. A spontaneous release of virtual photon pairs from the dressed vacuum can occur in a quantum optical system\cite{PhysRevLett.110.243601,PhysRevA.89.033827}. Three photon and multiphoton resonance can take place in the large-detuning Rabi model\cite{PhysRevA.92.023842,PhysRevA.102.053709}. Non intuitively, a photon can simultaneously excite two or more atoms\cite{PhysRevLett.117.043601,SR1,Bin:17,PhysRevA.95.063848,PhysRevA.96.063820,PhysRevA.101.053818}, which shows the reversal behavior of atomic two-photon or multiphoton absorbing processing. Parity-symmetry-breaking-induced cascade transition can significantly enhance the collective radiance in multi-atom Rabi model\cite{PhysRevA.99.033809} and N-phonon bundle emission is reported  in an acoustic cavity QED system recently\cite{PhysRevLett.124.053601}. What is more, nonclassical states such as entangled states could be prepared in USC regime\cite{PhysRevA.98.062327,PhysRevA.101.033809,SR2}.

\par The interaction between matters can be achieved not only by direct couplings, but also by indirect effective couplings\cite{nature1,PhysRevLett.85.2392,PhysRevLett.83.4204,PhysRevLett.87.037902,Tabuchi405,PhysRevA.95.063849,PhysRevA.96.043833,PhysRevA.96.023818}. A coherent coupling between a ferromagnetic magnon and a superconducting qubit through a microwave is experimentally confirmed\cite{Tabuchi405}. Specially separated qubits can be coupled by exchanges of virtual particles in cavity-QED is also reported both theoretically\cite{PhysRevLett.85.2392} and experimentally\cite{nature1,PhysRevLett.87.037902}. In 2019, Di Stefano \emph{et al} showed two mechanical oscillators in a cavity optomechanical system can coherently coupled together by exchanging of virtual photon pairs at a quantum level\cite{PhysRevLett.122.030402}. The model can also operate as a mechanical parametric down-converter and may find applications in the field of optomechanical quantum technology. Recently, Zhang \emph{et al} experimentally demonstrated that phonon heat is able to transfer across a vacuum through quantum fluctuations\cite{nature2}. Owning to the virtual excitation of the cavity, the coupling shows  robustness against cavity-induced loss and it has high application value in quantum information\cite{nature3}, nanotechnology devices\cite{nature2}, \emph{etc}. Consequently, is significant to find novel coherent couplings, trough the quantum vacuum, between kinds of matters.

\par Motivated by the above analysis, in this paper, we will show that in a hybrid system consisting of a two-photon Rabi model and a mechanical oscillator, a coupling between the atom and mechanical oscillator can be achieved by the exchange of virtual photon pairs induced by dynamical Casimir effect (DCE)\cite{PhysRevLett.122.030402,MooreDCE,nature4,PhysRevX}. This effective coupling is coherent and reversible in the absence of
losses. Interestingly, for the one-photon Rabi model, a Hawking-radiation-like behavior takes place\cite{HawkingRadiation}. The whole process can be described as follows. When there is a perturbation in the quantum vacuum by mechanical oscillator, a photon pair in the vacuum can be virtually excited. Owning to the strong coupling between cavity field and atom, one of the two virtual photons can be absorbed by the ground state atom, leaving the other simultaneously converted to real photon. During the whole process, the energy is conserved, e.g., $\omega_m=\omega_q+\omega_c$, where $\omega_m$ is the frequency of mechanical oscillator, $\omega_q$ is the atomic transition frequency, and $\omega_c$ is the resonant frequency of the cavity.

\par The remainder of this paper is organized as follows. In Sec.\ref{sec::2}, we introduce the two-photon Rabi model interacting with a single-mode mechanical oscillator through radiation pressure. The system Hamiltonian is numerically diagonalized and the resonant coupling between a phononic Fock state and atomic state is analyzed in Sec.\ref{sec::3}. The system dynamics is also shown by the master equation approach in USC regime\cite{PhysRevA.84.043832,PhysRevX,PhysRevLett.122.030402,PhysRevLett.109.193602,PhysRevA.92.023842,PhysRevA.89.033827,PhysRevA.91.013812,PhysRevA.98.053834} in Sec.\ref{sec::4}. In Sec.\ref{sec::5}, we discuss the one-photon Rabi model interacting with a mechanical oscillator and the Hawking-Radiation-like behavior is revealed. Experimental feasibility and potential applications are discussed in Sec.\ref{sec::6}. Finally, a conclusion is made in Sec.\ref{sec::7}.

\section{Model} \label{sec::2}
\begin{figure}
\includegraphics[scale=0.22]{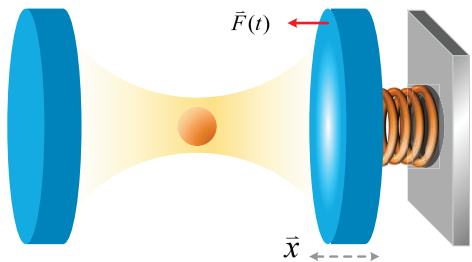}
\caption{A sketch of the Rabi model with one of the cavity mirrors connected by a mechanical oscillator, which can reach an oscillation frequency as high as $6$ GHz\cite{nature5}. We choose a microwave cavity here and its resonant frequency $\omega_c/2\pi\sim 5$ GHz. The atom in the Rabi model might be an artificial atom, e.g., a transmon qubit.}
\label{fig1}
\end{figure}
The model considered here is a hybrid system consisting of a two-photon Rabi model\cite{PhysRevA.97.013851} and a mechanical oscillator. As described in Fig.\ref{fig1}, one of the cavity mirrors is connected by a mechanical oscillator and oscillates at a frequency $\omega_m$ as high as $6$ GHz. This frequency region of mechanical oscillator has been realized by nanoresonator\cite{Rouxinol_2016}. The cavity is a single mode microwave cavity with a resonant frequency at about 5 GHz. The model can be simply regarded as a cavity optomechanical system with an atom inside. When the oscillation frequency gets the region of GHz, the DCE term can not be discarded\cite{PhysRevLett.112.203603,PhysRevA.93.022510}. By considering only one mechanical mode and the lowest cavity mode, the Hamiltonian of the whole system can be written as\cite{PhysRevA.51.2537,PhysRevLett.112.203603,PhysRevX} ($\hbar=1$)
\begin{equation}\label{e1}
H=H_{0}+V
\end{equation}
with
\begin{equation}\label{e2}
H_{0}=H_{q}+H_{om}
\end{equation}
and
\begin{equation}\label{e3}
V=\frac{1}{2}\kappa(a^{2}+a^{\dag2})(b+b^{\dag})+\lambda(a^{2}+a^{\dag2})\sigma_{x}.
\end{equation}
In Eq.\ref{e2}, $H_{q}=\omega_{q}\sigma_{z}/2$ is the Hamiltonian of bare atom, with $\omega_{q}$ and $\sigma_{z}$ being the atomic transition frequency and Pauli matrix, respectively. $H_{om}=\omega_{c}a^{\dag}a+\omega_{m}b^{\dag}b+\kappa a^{\dag}a(b+b^{\dag})$, with $\omega_{c}$ being the resonant frequency of the cavity, $a^{\dag}$ ($a$) being the creation (annihilation) operator of the cavity mode, $\omega_{m}$ being the frequency of the mechanical oscillator, $b^{\dag}$ ($b$) being the creation (annihilation) operator of the mechanical mode, and $\kappa$ being the coupling strength between cavity and mechanical oscillator, is the standard Hamiltonian of cavity optomechanical system. In Eq.\ref{e3}, the first term of $V$ describes the DCE and the second term, with $\sigma_{x}$ being the Pauli matrix, denotes the two-photon coupling between atom and cavity field with a coupling strength $\lambda$.

\par The standard optomechanics Hamiltonian $H_{om}$ can be analytically diagonalized by defining the displacement operators for the mechanical oscillator\cite{PhysRevX,PhysRevLett.122.030402,Liu_2018}, e.g., $\tilde{b}=b+\beta a^{\dag}a$ with $\beta=\kappa/\omega_{m}$. We can obtain
\begin{equation}\label{e4}
H_{om}=(\omega_{c}-\beta^{2}\omega_{m}a^{\dag}a)a^{\dag}a+\omega_{m}\tilde{b}^{\dag}\tilde{b}
\end{equation}
and the eigenvalue equation
\begin{equation}\label{e5}
 H_{om}|\varphi_{n,k}\rangle=E_{n,k}|\varphi_{n,k}\rangle,
\end{equation}
with
 \begin{equation}\label{e6}
E_{n,k}=n\omega_{c}-n^{2}\beta^{2}\omega_{m}+k\omega_{m}
 \end{equation}
and $|\varphi_{n,k}\rangle=|n\rangle\bigotimes|k_{n}\rangle$.
 Here, $|n\rangle$ is the bare photon state and $|k_{n}\rangle=D(n\beta)|k\rangle$, with $D(n\beta)=\textrm{exp}[n\beta(b-b^{\dag})]$ being the displacement operator and $|k\rangle$ being the bare mechanical oscillator Fock state, is the displaced mechanical oscillator Fock state. It indicates there are $k$ phonons in mechanical mode combined with $n$ photons in the cavity field. Apparently, $|k_{0}\rangle=|k\rangle$. By using Eq.\ref{e4} and Eq.\ref{e5}, $H_{0}$ in Eq.\ref{e1} can be easily diagonalized and the eigenstate can be defined as $|n, k_{n}, q\rangle$, with $|q\rangle=|g\rangle$ or $|e\rangle$ denoting the bare atomic state.
 \begin{figure}
\includegraphics[scale=0.44]{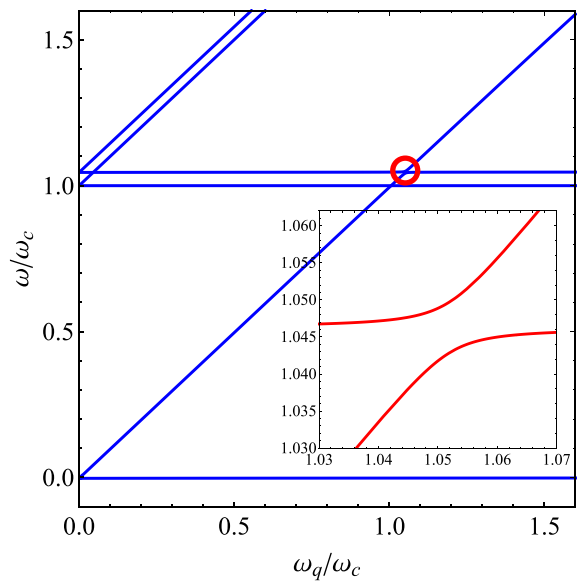}
\caption{Lowest-energy levels $\omega/\omega_{c}$ of the system Hamiltonian as a function of normalized atomic transition frequency $\omega_{q}/\omega_{c}$. In the plot, we assumed $\omega_{m}=1.05\omega_{c}$, $\kappa=\lambda=0.05\omega_{c}$. The inset is the enlarged view at the position of red circle. }
\label{fig2}
\end{figure}
\section{Rabi splitting} \label{sec::3}
We begin our study by numerically diagonalize the system Hamiltonian $H$ and indicate the resulting
energy eigenvalues and eigenstates as $\omega_{l}$ and $|\psi_{l}\rangle$ with $l=0,1,2,...$. We choose the labeling of states such that $\omega_{k}>\omega_{j}$ for $k>j$. The lowest-energy levels of $H$ as a function of normalized atomic transition frequency $\omega_{q}/\omega_{c}$ are shown in Fig.\ref{fig2}. In order to distinguish the energy levels contributed by phonon from that by photon, we have assumed there is a detuning between two frequencies, e.g., $\omega_{m}=1.05\omega_{c}$, in plotting Fig.\ref{fig2}. From Fig.\ref{fig2}, we observe that the energy level for $\omega/\omega_{c}\approx1.05$ (corresponding to state $|0,1,g\rangle$) and the energy level for $\omega/\omega_{c}\approx\omega_{q}$ (corresponding to state $|0,0,e\rangle$) shows an evident energy level anticrossing at $\omega_{q}/\omega_{c}\approx1.05$. The anticrossing originates from the coherent coupling between the one-phonon state $|0,1,g\rangle$ and the atomic excited state $|0,0,e\rangle$, with the cavity field invariably being the vacuum state. The physical process of the effective coupling can be interpreted as follows. When the mechanical oscillator is in its first excited state, it can transfer the energy to the vacuum of the cavity and excite two largely detuned virtual photons ($\sim1.05\omega_{c}/2$) by DCE. Owning to the strong coupling between atom and cavity field, the two photons can be absorbed simultaneously by the ground-state atom. That is to say, the mechanical oscillator can excite an atom trough the quantum vacuum. At the minimum energy splitting $2\Omega_{\textrm{eff}}\approx6.8\times10^{-3}\omega_{c}, \omega_{q}\approx1.052\omega_{c}$, with $\Omega_{\textrm{eff}}$ being the effective Rabi frequency, the two eigenstates of the system can be described by $|\psi_{2,3}\rangle\approx(1/\sqrt{2})\big(|0,1,g\rangle\pm|0,0,e\rangle\big)$, which are maximum entangled states between mechanical oscillator and atom. The effective coupling between states $|0,1,g\rangle$ and $|0,0,e\rangle$ can be analytically described by the effective interaction Hamiltonian\cite{PhysRevA.82.052106,PhysRevA.95.032124}
\begin{equation}\label{e7}
 H_{eff}=-\Omega_{\textrm{eff}}\big(|0,1,g\rangle\langle 0,0,e|+\textrm{H.c.}\big).
\end{equation}
\begin{figure}
\includegraphics[scale=0.32]{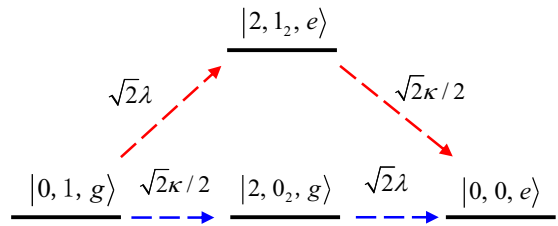}
\caption{A sketch describes two paths contributing to the effective coupling between state $|0,1,g\rangle$ and $|0,0,e\rangle$. There are two intermediate state $|2,0_{2},g\rangle$ and $|2,1_{2},e\rangle$ contribute to the transition. $\sqrt{2}\lambda$ and $\sqrt{2}\kappa/2$ are transition matrix elements.}
\label{fig3}
\end{figure}
\par The effective coupling can also be calculated by second-order perturbation theory. As displayed in Fig.\ref{fig3}, there are two paths connecting the transition $|0,1,g\rangle\leftrightarrow|0,0,e\rangle$. The main contribution is the path depicted by blue-dashed arrows. Although the intermediate state $|2,0_{2},g\rangle$ conserves the energy, the photon pair (Each photon has an energy at about $1.05\omega_{c}/2$) induced by DCE here is virtual owning to the large detuning to the cavity resonant frequency. The path connected by the intermediate state $|2,1_{2},e\rangle$ does not conserve the energy and the photon induced by DCE is also virtual. To apply the second-order perturbation theory, the basis vectors used here are the eigenstates of $H_{0}$ (see Eq.\ref{e1}) and the interaction Hamiltonian $V$ is regarded as a perturbation. With $|0,1,g\rangle$ being the initial state $|I\rangle$ and $|0,0,e\rangle$ being the final state $|F\rangle$, the effective coupling strength between $|0,1,g\rangle\leftrightarrow|0,0,e\rangle$ is given by\cite{PhysRevA.98.062327,PhysRevLett.122.030402,PhysRevX}
\begin{equation}\label{e8}
\begin{aligned}
&V_{\textrm{eff}}=-\Omega_{\textrm{eff}}=\sum_{l\neq I,F}\frac{V_{Fl}V_{lI}}{E_{I}-E_{l}}\\
&=\frac{\kappa\lambda}{\omega_{m}-2\omega_{c}+4\kappa^{2}/\omega_{m}}+\frac{\kappa\lambda}{-\omega_{q}-2\omega_{c}+4\kappa^{2}/\omega_{m}},
\end{aligned}
\end{equation}
where $V_{Fl}=\langle F|V|l\rangle$ is the transition matrix element, $|l\rangle$ is an eigenvector and $E_{l}$ is an eigenvalue of $H_{0}$. It is clearly shown that the Rabi frequency is proportional to the square of the coupling coefficient (assuming $\lambda=\kappa$). When $\lambda$ is small enough compared with the cavity frequency, the Rabi splitting vanishes. In order to get observable Rabi splitting, the coupling regime should be in SC or USC regime. In such regimes, the effective coupling becomes larger than the system decay rates, the transfer of one-phonon state to atomic excited state can be deterministic and reversible. If the system is initially prepared in the one-phonon state $|0,1,g\rangle$, neglecting decays, the system wave function will evolve as
\begin{equation}\label{e9}
|\psi(t)\rangle=\textrm{cos}(\Omega_{\textrm{eff}}t)|0,1,g\rangle-i\textrm{sin}(\Omega_{\textrm{eff}}t)|0,0,e\rangle.
\end{equation}
At time $t=\pi/\big(2\Omega_{\textrm{eff}}\big)$, the system will in the atomic excited state $|0,0,e\rangle$. While at time $t=\pi/\big(4\Omega_{\textrm{eff}}\big)$, the mechanical oscillator and atom will be in a maximally entangled state.
\begin{figure}
\includegraphics[scale=0.57]{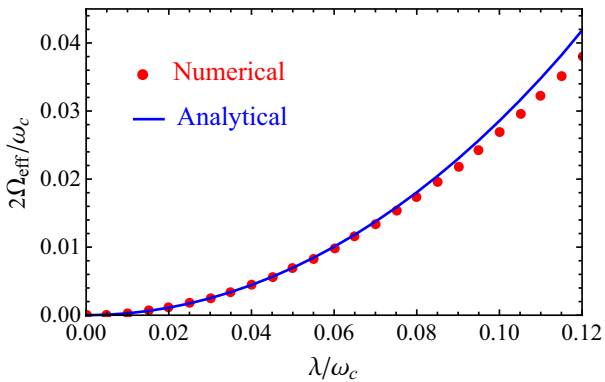}
\caption{Comparison between the numerically-calculated normalized Rabi splitting(red points) and the effective coupling calculated analytically by second-order perturbation theory (blue-solid line). In order to show the square-relation (not exactly) of coupling coefficient $\lambda$ or $\kappa$, we have assumed $\kappa=\lambda$. Other parameters are the same as in Fig.\ref{fig2}.}
\label{fig4}
\end{figure}
\par Fig.\ref{fig4} shows the comparison of the magnitudes of the normalized Rabi splitting $2\Omega_{\textrm{eff}}/\omega_{c}$ obtained analytically by second-order perturbation theory (see Eq.\ref{e7}) and that by numerically diagonalizing of the system Hamiltonian. We can see that the analytical result remains a very good approximation for normalized coupling strengths $\lambda/\omega_{c}\leq0.086$. When the value of $\lambda/\omega_{c}$ get larger, the effective coupling strength $V_{\textrm{eff}}$ becomes comparable to the unperturbed Hamiltonian $H_{0}$, which leads to the failure of the perturbation theory.

\section{System dynamics} \label{sec::4}
In order to better understand this effective coupling, a dynamic evolution process should be given. For describing a realistic system, all the dissipation channels need to be taken into account. Here, we adopt the master equation approach. However, for the coupling in USC regime, the standard quantum optical master equation breaks down\cite{PhysRevA.84.043832,PhysRevA.98.053834,PhysRevLett.109.193602}. Hence, we adopt a newly formed master equation in USC regime following Ref.\cite{PhysRevA.84.043832} and it can be written as follows
\begin{equation}\label{e10}
\begin{aligned}
\dot{\rho}(t)=&-i[H, \rho(t)]+\gamma_{a}\mathcal{D}[A^{-}]\rho(t)\\
&+\gamma_{m}\mathcal{D}[B^{-}]\rho(t)+\gamma_{q}\mathcal{D}[C^{-}]\rho(t).
\end{aligned}
\end{equation}
Here, $\rho(t)$ is the system density operator and constants $\gamma_{a},\gamma_{m},\gamma_{q}$ correspond to the damping rates for cavity mode, mechanical mode, and atom, respectively. The superoperator $\mathcal{D}$ is defined as
\begin{equation}\label{e11}
\mathcal{D}[O]\rho(t)=\frac{1}{2}\big[2O\rho(t)O^{\dag}-O^{\dag}O\rho(t)-\rho(t)O^{\dag}O\big],
\end{equation}
where $O=A^{-}, B^{-}$, or $C^{-}$ are dressed upping operators (negative frequency parts) for cavity field, mechanical mode, and atom, respectively. The dressed lowering operators (positive frequency parts) $O^{+}=O^{\dag}$ are defined as
\begin{equation}\label{e12}
O^{+}=\sum_{j,k>j}\langle\psi_{j}|(o+o^{\dag})|\psi_{k}\rangle|\psi_{j}\rangle\ \! \langle\psi_{k}|,
\end{equation}
with the bare operators $o=a, b ,\sigma_{-}$. In the process of deriving the master equation (Eq.\ref{e10}), the Markov approximation is applied and the temperature of the reservior is assumed to be zero. It is demonstrated that influence of temperature on the mechanical expectation values is almost negligible\cite{PhysRevLett.122.030402}.
\begin{figure}
\includegraphics[scale=0.54]{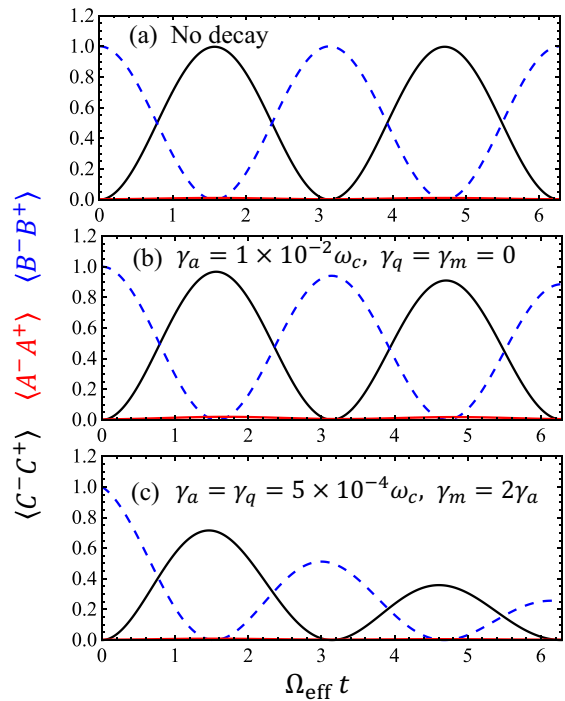}
\caption{Time evolution of the atomic mean excitation number $\langle C^{-}C^{+}\rangle$ (black-solid line), the cavity mean photon number $\langle A^{-}A^{+}\rangle$ (red-dotted line), and the mean phonon number $\langle B^{-}B^{+}\rangle$ (blue-dashed line), assuming the system is initially in the one-phonon state $|0,1,g\rangle$. (a) System dynamics with no decay. (b) System dynamics with ultra-high cavity decay rate $\gamma_{a}=1\times10^{-2}\omega_{c}, \gamma_{m}=\gamma_{q}=0$. (c) System dynamics with decay rates $\gamma_{a}=\gamma_{q}=5\times10^{-4}\omega_{c}, \gamma_{m}=2\gamma_{a}$. Other parameters are the same as in Fig.\ref{fig1}.}
\label{fig5}
\end{figure}
\par We start our analysis by numerically solving the master equation (Eq.\ref{e10}) in a truncated M-dimension Hilbert space spanned by the eigenvectors of the system Hamiltonian $H$. The truncation is valid only when increasing the dimension number M, the system dynamics is not significantly affected \cite{PhysRevA.92.023842,PhysRevX}. When solving the master equation, the atomic transition frequency is fixed ($\omega_{q}/\omega_{c}\approx1.052$) where the minimum energy splitting takes place.
\par We assume the system is initially in the one-phonon state $|0,1,g\rangle$, time evolution of the atomic mean excitation number $\langle C^{-}C^{+}\rangle$ (black-solid line), the cavity mean photon number $\langle A^{-}A^{+}\rangle$ (red-dotted line), and the mean phonon number $\langle B^{-}B^{+}\rangle$ (blue-dashed line) is shown in Fig.\ref{fig5}. Fig.\ref{fig5}(a) displays the ideal case that all the decay rates are zero. In such a case, the mean phonon number varies at an initial value \emph{1} and shows a cosinoidal evolution with time at a frequency $\Omega_{\textrm{eff}}$. Meanwhile, the atomic mean excitation number varies at an initial value \emph{0} and shows a sinusoidal evolution with time at the same frequency. At a time $t=\pi/\big(2\Omega_{\textrm{eff}}\big)$, the mean phonon number gets \emph{0} and the atom jumps to its excited state. This process is reversible and reveals an energy exchange between the mechanical oscillator and the atom. It is consistent with the above analysis (see Eq.\ref{e9}). It is worth noting that the mean photon number is almost zero during the whole process. Owning to DCE, we also observe that when the mechanical oscillator is in its vacuum state (at the time $\Omega_{\textrm{eff}}t=(n+1/2)\pi, n=0,1,2,...$), the mean photon number is not exact zero and gets its maximum. Fig.\ref{fig5}(b) displays the case that the cavity has an ultra-low quality factor $Q_{c}=\omega_{c}/\gamma_{c}=100$ and other decay rates are zero. Although the cavity decay rate is ultra high, the Rabi oscillating behavior of the atom and mechanical oscillator is almost unaffected. That is to say, the effective coupling is robust against cavity lose. When taking all the decay channels into account, as shown in Fig.\ref{fig5}(c), the mean values still oscillate in cosine (or sine) form, but their amplitudes decrease exponentially, as expected.
\begin{figure}
\includegraphics[scale=0.42]{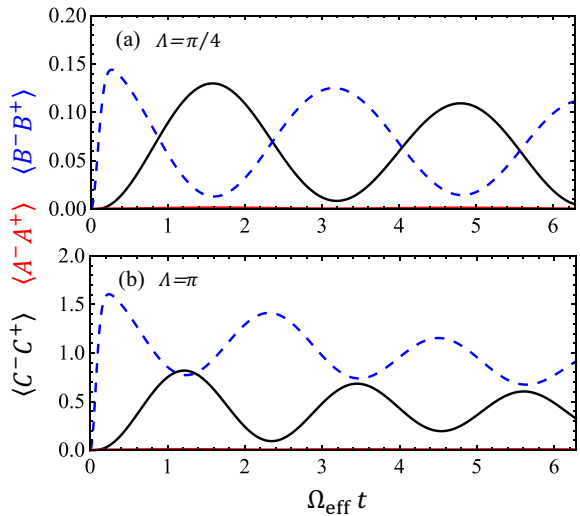}
\caption{Time evolution of the atomic mean excitation number $\langle C^{-}C^{+}\rangle$ (black-solid line), the cavity mean photon number $\langle A^{-}A^{+}\rangle$ (red-dotted line), and the mean phonon number $\langle B^{-}B^{+}\rangle$ (blue-dashed line) after the arriving of a $\pi$-like Gaussian pulse. The system is initially prepared in its ground state $|\psi_{0}\rangle\approx|0,0,g\rangle$. (a) System dynamics with $\Lambda=\pi/4$. (b) System dynamics with $\Lambda=\pi$. The decay rates are $\gamma_{a}=\gamma_{q}=\gamma_{m}=1\times10^{-4}\omega_{c}$. Other parameters are the same as in Fig.\ref{fig1}. }
\label{fig6}
\end{figure}
\par In order to simulate a more realistic scene, we first prepare the system in its ground state $|\psi_{0}\rangle\approx|0,0,g\rangle$ and drive the mechanical oscillator with a $\pi$-like Gaussian pulse. The drive Hamiltonian is given by\cite{PhysRevX,PhysRevLett.122.030402}
\begin{equation}\label{e13}
H_{d}=\mathcal{F}(t)\big(e^{i\omega_{d}t}B^{+}+\textrm{H.c.}\big),
\end{equation}
where $\mathcal{F}(t)=\Lambda\mathcal{G}(t-t_{0})$, with $\Lambda$ being the amplitude and $\mathcal{G}(t-t_{0})$ being the normalized Gaussian function. We choose the frequency of the drive $\omega_{d}=\omega_{m}$ and the standard deviation of the Gaussian function $\sigma=1/\big(10\Omega_{\textrm{eff}}\big)$ in plotting Fig.\ref{fig6}. System decay channels are all taken into account. Fig.\ref{fig6} displays the system dynamics after the arriving of a $\pi$-like Gaussian pulse. It is clearly shown that the mean phonon number $\langle B^{-}B^{+}\rangle$ starts from $0$ and grows rapidly when the pulse comes. After $\langle B^{-}B^{+}\rangle$ getting its maximum, the atomic mean excitation number $\langle C^{-}C^{+}\rangle$ starts to grow and gets its maximum when $\langle B^{-}B^{+}\rangle$ gets its minimum. The two quantities varies sinusoidally with peak amplitudes decaying exponentially. The amplitude $\Lambda$ significantly affects the maximum value of $\langle B^{-}B^{+}\rangle$. We also observed that the mean photon number is almost $0$ during the whole process, which once again demonstrates the energy exchange between mechanical oscillator and atom is accomplished by the exchange of \emph{virtual} photons.

\section{One-photon process: a phenomenon similar to Hawking radiation} \label{sec::5}
Section \ref{sec::2} describes the model of a two-photon Rabi model interacting with a single-mode mechanical oscillator, where two virtual photons induced by DCE can be jointly absorbed by the atom. One may wonder when considering the one-photon Rabi model, where the atom can only absorb one of the two virtual photons, what will happen to the unabsorbed photon. In this section, we try to answer this question.
The one-photon Rabi model is the standard Rabi model\cite{Rabimodel}, which describes that the atom can only absorb or emit one photon during the Rabi oscillation. When the standard Rabi model with one of its cavity mirrors connected by a single mode mechanical oscillator, the system Hamiltonian has the same from as the two-photon case (see Eq.\ref{e1} and \ref{e2}), but the interaction Hamiltonian $V$ in Eq.\ref{e3} should be modified as
\begin{equation}\label{e14}
V=\frac{1}{2}\kappa(a^{2}+a^{\dag2})(b+b^{\dag})+\lambda(a+a^{\dag})\sigma_{x}.
\end{equation}
The first term in Eq.\ref{e14} is still the DCE term and the last term describes the one-photon process between atom and cavity field.
\par
\begin{figure}
\includegraphics[scale=0.44]{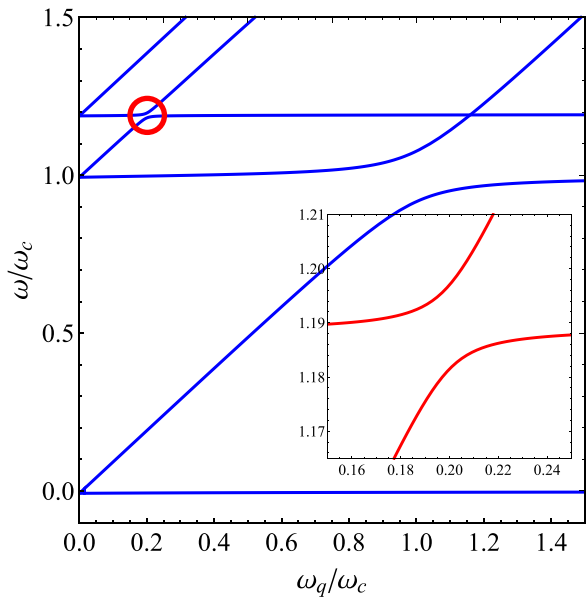}
\caption{Lowest-energy levels $\omega/\omega_{c}$ of the system Hamiltonian (one-photon case) as a function of normalized atomic transition frequency $\omega_{q}/\omega_{c}$. In the plot, we assumed $\omega_{m}=1.2\omega_{c}$, $\kappa=\lambda=0.08\omega_{c}$. The inset is the enlarged view at the position of red circle.}
\label{fig7}
\end{figure}

 Following the method applied in Sec.\ref{sec::3}, we first numerically diagonalize the system Hamiltonian. The lowest-energy levels $\omega/\omega_{c}$ of the system Hamiltonian (one-photon process) as a function of normalized atomic transition frequency $\omega_{q}/\omega_{c}$ are shown in Fig.\ref{fig7}. Here, we have assumed $\omega_{m}=1.2\omega_{c}$, $\kappa=\lambda=0.08\omega_{c}$. If the cavity is a microwave cavity with a resonant frequency $\omega_{c}\sim 5 \textrm{GHz}$, frequency of the mechanical oscillator $\omega_{m}=1.2\omega_{c}\sim 6 \textrm{GHz}$ is experimentally available\cite{nature5}. Fig.\ref{fig7} clearly shows that there is an evident anticrossing between the energy level for $\omega/\omega_{c}\approx1.2$ (corresponding to state $|0,1,g\rangle$) and the energy level for $\omega/\omega_{c}\approx1+\omega_{q}/\omega_{c}$ (corresponding to state $|1,0_{1},e\rangle$) at $\omega_{q}/\omega_{c}\approx0.2$. This energy level avoided-crossing originates from the coherent coupling between the one-phonon state $|0,1,g\rangle$ and the atomic excited state accompanied by a dressed one-photon state. At the minimum energy splitting $2\Omega_{\textrm{eff}}\approx1.55\times10^{-2}\omega_{c}, \omega_{q}\approx0.199\omega_{c}$, the two eigenstates of the system can be approximated as $|\psi^{\prime}_{3,4}\rangle\approx(1/\sqrt{2})\big(|0,1,g\rangle\pm|1,0_{1},e\rangle\big)$, where the superscript stands for the quantities in one-photon case. $|\psi^{\prime}_{3,4}\rangle$ are Greenberger-Horne-Zeilinger (GHZ) states for cavity field, mechanical oscillator, and atom.
\par
\begin{figure}
\includegraphics[scale=0.44]{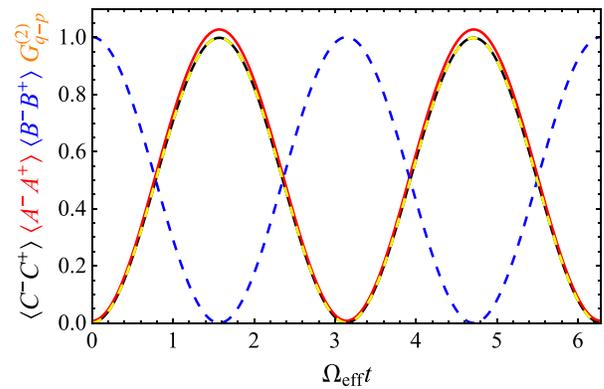}
\caption{Time evolution (one-photon case) of the atomic mean excitation number $\langle C^{-}C^{+}\rangle$ (black-solid line), the cavity mean photon number $\langle A^{-}A^{+}\rangle$ (red-solid line), the mean phonon number $\langle B^{-}B^{+}\rangle$ (blue-dashed line), and the zero-delay second-order correlation function of atom and photon $G^{(2)}_{q-p}$ (yellow-dashed line), assuming the system is initially in the one-phonon state $|0,1,g\rangle$. We are only interested in the ideal case such that $\gamma_{a}=\gamma_{q}=\gamma_{m}=0$. Other parameters are the same as in Fig.\ref{fig7}.}
\label{fig8}
\end{figure}

 In order to fully understand this anomalous avoided crossing, system dynamics should be presented. The atomic mean excitation number $\langle C^{-}C^{+}\rangle$ (black-solid line), the cavity mean photon number $\langle A^{-}A^{+}\rangle$ (red-solid line), the mean phonon number $\langle B^{-}B^{+}\rangle$ (blue-dashed line), and the correlation function of atom and photon $G^{(2)}_{q-p}=\langle C^{-}A^{-}A^{+}C^{+}\rangle$ (yellow-dashed line) vary with time $\Omega_{\textrm{eff}}t$ are shown in Fig.\ref{fig8}. We are only interested in the ideal case and the system assumed to be initially in the one-phonon state $|0,1,g\rangle$. Dynamics of the atomic mean excitation number and mean phonon number shows exact coincidence with that in Fig.\ref{fig5}(a). Yet, the mean photon number is no longer zero but varies sinusoidally with an amplitude approaching one, keeping step with the atomic mean excitation number. This means the photon is no longer virtual but a real photon. From Fig.\ref{fig8}, we can also observe that the atomic mean excitation number $\langle C^{-}C^{+}\rangle$ and the correlation function $G^{(2)}_{q-p}$ almost coincide at any time. This indicates the real photon conversion and atomic excitation is almost simultaneous\cite{PhysRevLett.117.043601}. Fig.\ref{fig8} also indicates the energy exchange between mechanical oscillator, cavity field, and atom. In the first-half Rabi period, the energy of mechanical oscillator transfers to the atom and cavity field. In the second-half Rabi period, the opposite is true. During the whole process, the energy is conserved, $\omega_{m}=\omega_{q}+\omega_{c}$.
\par These results can be interpreted as follows. When there is a perturbation by mechanical oscillator to the vacuum field of the cavity, a photon pair can be induced and partially excited by DCE. If the DCE resonant criterion $k\omega_{m}=2\omega_{c}, k=1,2,...$\cite{PhysRevX} is not satisfied, the two photons are virtual. However, when there is a resonantly-strong-coupled ground-state atom in the vacuum, one of the virtual photons can be immediately absorbed by the atom. The other photon, with no virtual photon to pair, simutaneously converts to a real one. This phenomenon shows similarity to Hawking radiation. During the whole process, the ground state atom plays the role of a black hole and the real photon acts as the radiation emitted by the black hole.

\section{Experimental feasibility and potential applications} \label{sec::6}
For experiment, one can set up the experimental devices as shown in Fig.\ref{fig1}. The atom used here should be a transmon qubit in an ultrastrongly-coupled circuit-QED system and the cavity optomechanical system should be ultrahigh-frequency mechanical micro- or nanoresonators, whose oscillating frequency is in the GHz spectral range\cite{nature5,Rouxinol_2016}. We take the one-photon case for instance. In order to observe such a resonant coupling, we can first prepared the system in the uncoupled ground state $|0,0,g\rangle$. Then we drive the mechanical oscillator by a $\pi$-like Gaussian pulse ($\omega_{d}=\omega_{m}=1.2\omega_{c}$) and detect the cavity field. There is no Casimir photon owning to the large detuning. Next, we adiabatically tune the qubit frequency into resonance with the virtual Casimir photon $\omega_{q}=\omega_{m}-\omega_{c}\approx0.199\omega_{c}$ and detect the cavity field and atomic state at time  $t=\pi/\big(2\Omega_{\textrm{eff}}\big)$. If the cavity is in the one-photon state and the atom is in its excited state, our theoretical prediction can then be demonstrated. The parameters used in the paper are experimental feasible and the predicted results can hence be observed in present day laboratories\cite{nature1,RevModPhys.84.1}.
\par Our theoretical proposal can find application in the field of quantum information, e.g., entangled state creation. For the two-photon case, a Bell state for mechanical oscillator and atom can be prepared at time $t=\pi/\big(4\Omega_{\textrm{eff}}\big)$. Although the proposal of creation Bell state was put forward in ultrastrongly-coupled circuit-QED system in Ref.\cite{PhysRevA.98.062327}, our method has a clear advantage owning to its robustness against cavity lose. For the one-photon case, a GHZ state for atom, mechanical oscillator, and the cavity filed can also be obtained.
\par In the one-photon case, the creation of real photon, accompanied by exciting an atom from ground to exited state, from the quantum vacuum can also be used to interpret and simulate the well-known Hawking radiation. Moreover, our proposal may find application in the field of nanotechnology such as on-chip heat transfer\cite{nature2}.

\section{Conclusions} \label{sec::7}
In conclusion, we have studied the Rabi model with one of the cavity mirrors connected by a mechanical oscillator. For the two-photon case, an effective coupling between the mechanical oscillator and atom, with the cavity field invariably being the vacuum state, can be achieved by the exchange of virtual photon pairs induced by DCE. The vacuum mediated effective coupling is robust against cavity lose. For the one-photon case, a Hawking-radiation-like phenomenon occurs: the ground state atom can absorb a photon from a virtual photon pair and leave the other photon converted to a real one. Our theoretical proposal is experimentally visible in present day laboratories and may find applications in the field of quantum information, nanotechnology and so on.
\begin{acknowledgments}
The author would like to thank Bo Wang for helpful discussions and Vincenzo Macr\`{\i} for technical support. This work was supported by the National Natural Science Foundation of China (No. 11204088), Science and Technology Program of Guangzhou, China (No.201607010019), and the Fundamental Research Funds for the Central Universities.
\end{acknowledgments}

\bibliographystyle{apsrev4-2}

\end{document}